%% file: paper.tex
\documentclass[aps,prd,showpacs,preprintnumbers]{revtex4}
\usepackage{amsmath,amssymb}
\usepackage{graphicx}

\input{macros.tex}

\begin{document}
\title{Bayesian priors and nuisance parameters}
\author{Sourendu\ \surname{Gupta}}
\email{sgupta@theory.tifr.res.in}
\affiliation{Department of Theoretical Physics, Tata Institute of Fundamental
         Research,\\ Homi Bhabha Road, Mumbai 400005, India.}
\author{Anirban\ \surname{Lahiri}}
\email{alahiri@physik.uni-bielefeld.de}
\affiliation{Fakult\"at f\"ur Physik, Universit\"at Bielefeld,\\
         D-33615 Bielefeld, Germany.}
\begin{abstract}
Bayesian techniques are widely used to obtain spectral functions from
correlators.  We suggest a technique to rid the results of nuisance
parameters, \ie, parameters which are needed for the regularization but
cannot be determined from data.  We give examples where the method works,
including a pion mass extraction with two flavours of staggered quarks
at a lattice spacing of about 0.07 fm. We also give an example where
the method does not work.
\end{abstract}

\maketitle

Some years ago, Bayesian techniques were introduced for the analysis
of correlation functions obtained in lattice gauge theory at zero
\cite{bayes} and finite \cite{tikhonov,lgtmem} temperatures. Although
the details required from these analyses differ, there are many
similarities. In both cases one needs to analyze measurements of
correlation functions, $C(t)$, where $t$ is the separation between
the source and sink operators in Euclidean time. The analysis extracts
some features of a spectral function, $\rho(\omega)$, where $\omega$
is an ``energy'' variable.  At zero temperature $\rho(\omega)$ is a sum
of delta functions, and the locations and strengths of the ones with
smallest $\omega$ are interesting.  At finite temperature the shape
of $\rho(\omega)$ at low $\omega$ is the object of interest.  In both
cases, the number of features in $\rho(\omega)$ far exceed the number of
measurements, and one needs to isolate the features of interest. One
may look upon Bayesian techniques as regularization techniques.
Bayesian techniques allow us to introduce assumptions about aspects of
$\rho(\omega)$, and then to relax some of these assumptions as dictated
by data. The procedure involves extraneous (nuisance) parameters, having
no physical interpretation, whose treatment remains an open question
\cite{guber}. We address the treatment of these parameters here.

Let us begin with a linear fitting problem, such as the one which
arises in the extraction of the spectral function at finite temperature
\cite{tikhonov}. Given a kernel $K$ and the model
\beq
   C(t) = K_{t i}\rho_i, \qquad
    (1\le t\le N_t,\;{\rm and}\;1\le i\le N_e),
\eeq{model}
we would like to find the optimum values of $\rho_i$ which describe a set
of data $C(t)$ with covariance matrix $\Sigma$. If the data are scaled by
a factor $\lambda$, then the $\rho_i$ are also scaled by the same factor.

The familiar case is $N_t>N_e$. A simple example is to fit a straight
line to three points. As we know, are no solutions unless there are
linear combinations of the data which do not determine the parameters,
as happens if some of the data points are related to each other by
adding vectors in the null-space of $K$.  If $\Sigma\ne0$ then one can
obtain a consistent solution, even otherwise, by minimizing the norm of
$C-K\rho$ using $\Sigma^{-1}$ as a metric.  This is the familiar case;
the function to be minimized is
\beq
   \chi^2(\rho) = \frac12 \Delta^T(\rho)\Sigma^{-1}\Delta(\rho),
     \qquad{\rm with}\qquad \Delta(\rho)=C-K\rho.
\eeq{maxl}
In the space of the parameters $\rho$, the function $\chi^2$ is quadratic with
no flat directions. Minimizing this corresponds to maximizing the conditional
probability of the measurements $C$ given the spectral function $\rho$,
\beq
   P(C|\rho) \propto \exp\left[-\frac12 \Delta^T(\rho)\Sigma^{-1}\Delta(\rho)\right].
\eeq{gaussian}
The probability arguments are all familiar for Gaussian distributed
measurements \cite{lyons}.

When $N_t<N_e$ these methods are ill-defined because the data does not
constrain the parameters. In other words, the function $\chi^2$ has many
flat directions.  The problem can be regularized if one has a prior
guess $\rho^{(0)}$. One includes this by maximizing a probability obtained
by using Bayes' theorem---
\beq
   \frac{\partial P(\rho|C,\rho^{(0)},\alpha)}{\partial\rho_i} =0, \qquad{\rm where}\qquad
   P(\rho|C,\rho^{(0)},\alpha) = \frac{P(C|\rho) P(\rho|\rho^{(0)},\alpha)}{P(C|\rho^{(0)},\alpha)}.
\eeq{maxim}
for $P(C|\rho)$ and $P(C|\rho^{(0)},\alpha)$ one uses the expression of eq.\
(\ref{gaussian}).  The maximum entropy model (MEM) for the new factor is
\beq
   P(\rho|\rho^{(0)},\alpha) \propto \exp\left(\alpha\sum_{i=1}^{N_e}\left[\rho_i-\rho^{(0)}_i-\rho_i
      \log\left(\frac{\rho_i}{\rho^{(0)}_i}\right)\right]\right).
\eeq{mem}
The parameter $\alpha$ is a nuisance parameter.
We can write a real function $f(\rho) =-\log P(\rho|C,\rho^{(0)},\alpha)$, so that the
minimization condition reduces to
\beq
   \frac{\partial f(\rho)}{\partial\rho_i} = -K_{\mu i}\Sigma^{-1}_{\mu\nu}
     \Delta_\nu(\rho) + \alpha\log\left(\frac{\rho_i}{\rho^{(0)}_i}\right) = 0.
\eeq{soln}
This description of the MEM follows \cite{bryan}. Similarly, the method of
constrained fit (MCF) \cite{bayes} can be defined by
\beq
   P(\rho|\rho^{(0)},\sigma^{(0)}) \propto \exp\left[-\frac12\sum_{i=1}^{N_e}
    \left(\frac{\rho_i-\rho^{(0)}_i}{\sigma^{(0)}_i}\right)^2\right].
\eeq{mcf}
Here the parameters $\sigma^{(0)}_i$ are nuisance parameters.

There are several suggestions in the literature on how to treat the MEM
nuisance parameter $\alpha$. Note that if we had immense confidence in
the prior, then we would replace Bayes' theorem in \eqn{maxim} by an
equation which treats $\alpha$ as a Lagrange multiplier implementing the
constraint. However, it is precisely when the prior is not a certainty
that interesting things happen, and we allow the data to change our state
of knowledge \cite{jaynes}.  Changing $\alpha$ in \eqn{mem} corresponds
to changing the width of $P(\rho|\rho^{(0)})$ around the modal value
$\rho^{(0)}$.  One suggestion, such as in the first of \cite{bayes}, is
to construct the posterior distribution $P(\alpha|C)$ and then use the
most probable $\alpha$.  An alternative was suggested in \cite{guber} that
one averages over all values of $\alpha$ with weight $P(\alpha|C)$. This
was the procedure followed in \cite{lgtmem,burnier}. Although this is
consistent with statistical practice for unobserved parameters, the
question is whether the degree of confidence in the prior is a normal
statistical parameter.  Here we examine a different approach to removing
the nuisance parameters.

Before proceeding it is worthwhile to consider the nomenclature
of nuisance parameters.  We have used the term in the sense that
these are parameters which cannot be extracted from the measurements
at hand. The parameters $\alpha$ in the MEM, or $\sigma^{(0)}_i$ in
the MCF, quantify confidence limits on the priors. In applications to
lattice QCD, they are entirely qualitative, and therefore a nuisance,
not to be trusted. However, if these techniques are used to combine
multiple data sets, then they are no longer nuisance parameters,
but can carry important information. In applications such as detector
characterization \cite{detector} there is no need to remove parameters
such as these. One may imagine also that global fits where different
experimental data sets are combined, such as neutrino oscillation
parameters, CKM matrix analysis, or structure function extractions,
could adopt Bayesian treatments where these parameters are used to check
consistency of data sets.

\bef
\begin{center}
\includegraphics[scale=0.7]{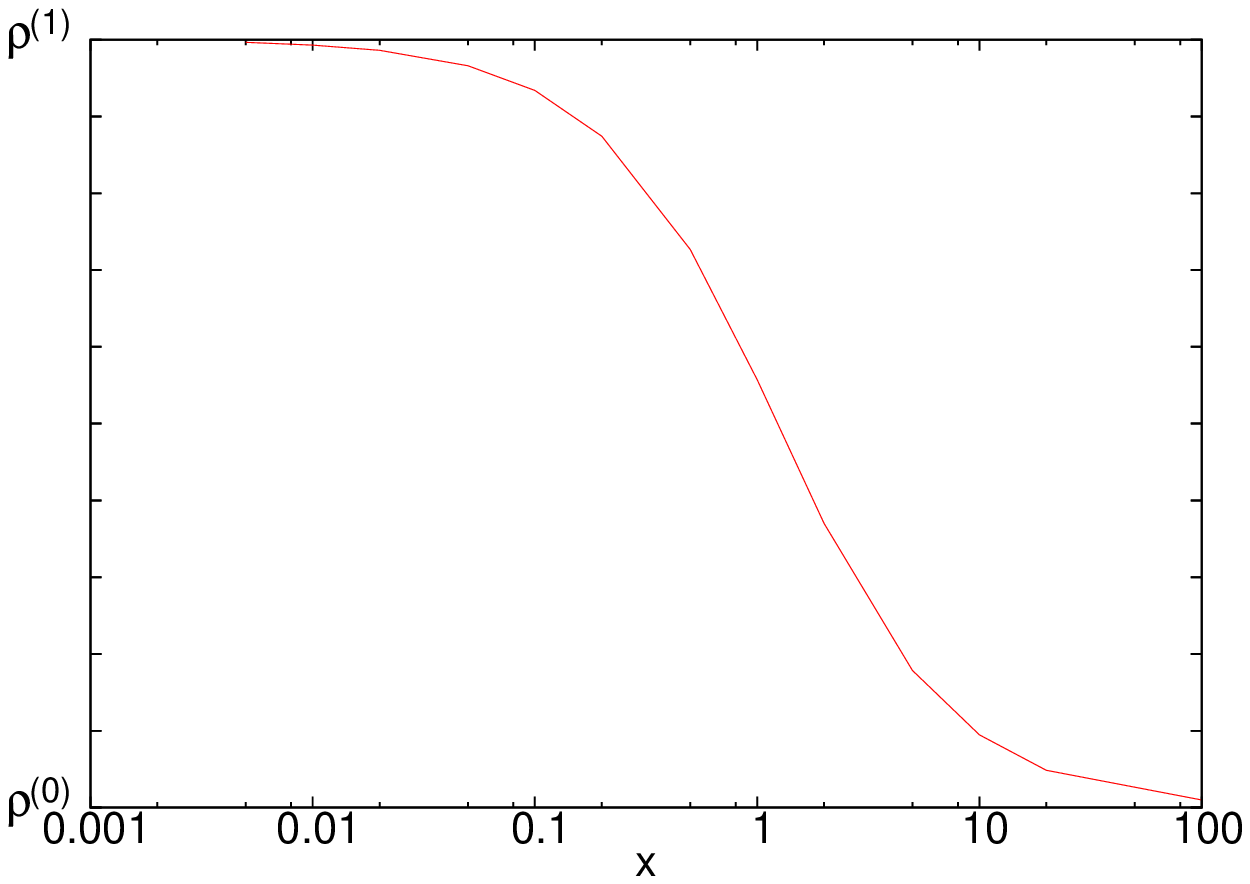}
\end{center}
\caption{The solution of \eqn{simpler}, as a function of $x$, shows a cross
 over from a region essentially determined by the data, \ie, $\rho=\rho^{(1)}$,
 to one where it is determined by the prior, \ie, $\rho=\rho^{(0)}$.}
\eef{crossover}

Maximum likelihood, \ie, minimizing the $\chi^2$ of \eqn{maxl} works
when the $N_t>N_e$. However, nothing prevents us from including a prior
into the fit, especially when one is known from other experiments. We
develop intuition by understanding this simple case.  Take the simplest
example, that of fitting a constant to a set of data, so that $N_e=1$. A
practical application is of extracting a mass from an observation of a
flat plateau in local masses. Assume that $\Sigma$ is diagonal. Define
\beq
   d_2=\sum_{t=1}^{N_t} \frac{C_t^2}{\sigma_t^2},\qquad
   d_1=\sum_{\mu=1}^{N_t} \frac{C_t}{\sigma_t^2},\qquad
   d_0=\sum_{\mu=1}^{N_t} \frac1{\sigma_t^2}.
\eeq{defs}
In this example we will think of $C_t$ as local masses, $\sigma_t$ as
their errors, and $\rho$ to be the estimate of the mass. In the MEM
framework, the function to be minimized is
\beq
   f = \frac{d_0}2\left[\rho^2-2\,\frac{d_1}{d_0}\rho+\frac{d_2}{d_0}\right]
     - \alpha\left[\rho-\rho^{(0)}-\rho\log\left(\frac\rho{\rho^{(0)}}\right)\right].
\eeq{func}
Introduce the notation $\rho^{(1)}=d_1/d_0$ and $x=\alpha/d_0$. The minimum
occurs at the solution of
\beq
   \rho-\rho^{(1)}=-x\log\left(\frac\rho{\rho^{(0)}}\right).
\eeq{simpler}
At $x=0$ the prior plays no role, and the solution is the usual
maximum likelihood solution $\rho=\rho^{(1)}$.  The right hand
side vanishes at the prior $\rho=\rho^{(0)}$, and this is the
solution when $x\to\infty$.  For $x\ll1$ the solutions remains
close to $\rho^{(1)}$ and for $x\gg1$ it is close to $\rho^{(0)}$,
with a cross over between the two regimes occurring when $x\simeq1$,
as shown in \fgn{crossover}.
When $\alpha=0$ the statistical error, $\sigma$, on the fitted parameter
is obtained by solving $f(\rho^{(1)}+\sigma)=f(\rho^{(1)})+1$. This gives
$\sigma=\sqrt{2/d_0}$. When the prior is included, one could still define
the error by the criterion that $f(\rho)$ changes by unity when $\rho$
is changed around the minimum by an amount $\sigma$.  One finds that
$\sigma$ decreases as $x$ increases, and, in the limit of large $x$,
the error $\sigma$ is controlled by the value of $x$.  The best fit
value and the error then show that the MEM prior determines the fit
for large $\alpha$ and the data determines the fit for small $\alpha$,
with a cross over between the two regimes when $\alpha\simeq d_0$. One
expects a similar crossover between prior-driven and data-driven regimes
also in MCF.

It was observed recently that the distribution of correlation functions
measured in lattice simulations are highly non-Gaussian \cite{ilgti}.
However, it was also shown in  that it is possible to analyze this data
using modern simulation methods coupled to appropriate bootstrap analysis
in such a way that the distribution of local masses, for example, is
close enough to Gaussian that the arguments above continue to hold. Even
in a non-Gaussian analysis, one may use \eqn{soln} or its equivalent
as a regularization, without using the probability interpretation in
\eqn{gaussian} or \eqn{maxim}.

We explore these methods using the data set used in \cite{ilgti}.
Correlators were measured in two-flavour QCD using staggered quarks. The
statistical analysis of the Goldstone pion correlators and local masses
are discussed in detail in \cite{ilgti}.  Here we discuss the extraction
of a mass from the measured local masses.  The quantities to be fitted are
a tower of masses, $m_0<m_1<m_2\cdots$ and the corresponding amplitudes
$A_i$. We chose to analyze the data for $\beta=5.7$ and bare quark
mass $ma=0.025$ because it was reported in \cite{ilgti} that the local
mass plateau could was not well developed for this parameter set. This
corresponds to a lattice spacing of about 0.07 fm when the scale is set
by the Wilson flow scale $w_0$. As a result, there was a minor mismatch
between the fitted mass and the local masses. It is interesting to do
a Bayesian analysis to see whether the lowest mass can be obtained with
better confidence.

\bef
\begin{center}
\includegraphics[scale=0.7]{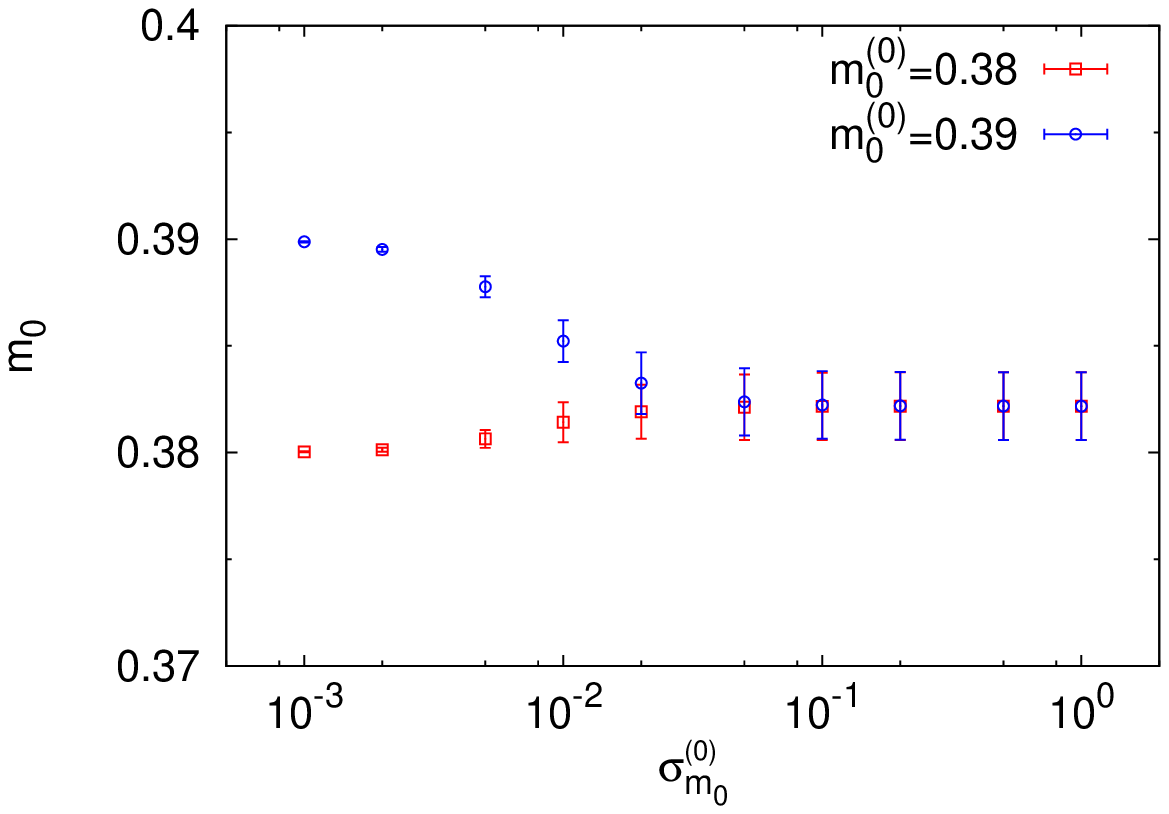}
\includegraphics[scale=0.7]{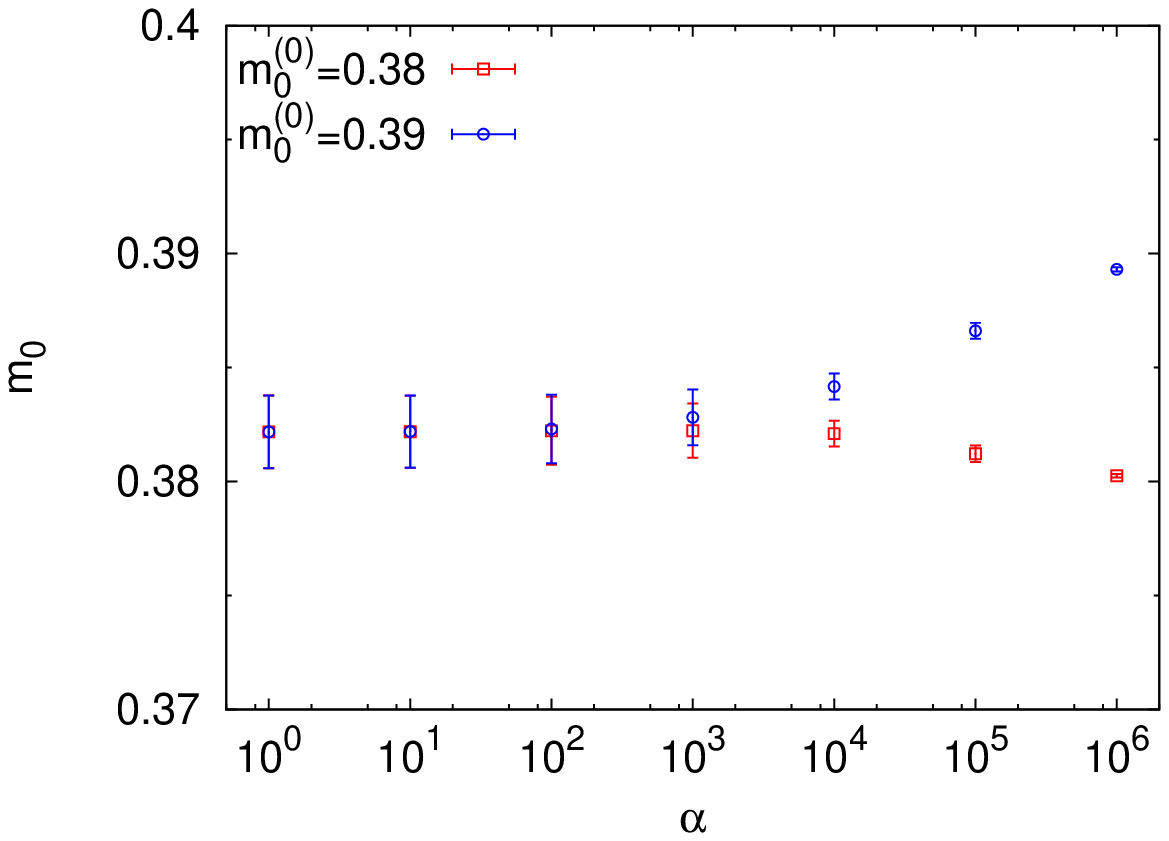}
\end{center}
\caption{(First panel) The best fit value of $m_0$ changes systematically
 with the nuisance parameter $\sigma^{(0)}_{m_0}$ in MCF. For small values
 of this parameter the best fit to the mass is determined largely by the
 prior.  However, for larger values, the best fit does not depend on
 $m_0^{(0)}$. (Second panel) The dependence on the nuisance parameter
 $\alpha$ in MEM is the inverse. Both these figures are for the analysis
 of pion masses at $\beta=5.7$ and bare quark mass $ma=0.025$.}
\eef{stable}

In the MCF we write the prior values $m_i^{(0)}$, $A_i^{(0)}$, and the
nuisance parameters $\sigma^{(0)}_{m_i}$ and $\sigma^{(0)}_{A_i}$. In
the MEM there is a single nuisance parameter $\alpha$. One sees from
\eqn{mem} and \eqn{mcf} that the role played by $\alpha$ is similar to
$1/\sigma^{(0)}_p$ where $p$ stands for any parameter.  We argued above
in a particularly simple model that there is a cross over between two
kinds of behaviour on tuning $\alpha$: for large $\alpha$ the best fit
and its error is determined essentially by the prior, for small $\alpha$
the data determines the fit. In \fgn{stable} we show that this behaviour
is obtained also for a realistic fit.

A similar procedure can be followed for those $m_i$ which cannot be
fixed by the measurements, and are therefore also nuisance parameters
in the same sense as above.  For all these parameters we test whether
the results for the remaining parameters are stable under change of all
the nuisance parameters.  Although in principle there is an infinite
tower of masses, in practice we found that adding two or three masses
was sufficient to give a stable fit for $m_0$.  We found that $m_0$
is measurable, but other masses are not.  As a result, the extraction
of the ground state mass corresponds to $N_t>N_e$. Bayesian methods
gave an estimate of the mass
\beq
  am_0 = 0.382\pm0.002  \qquad\qquad(\beta=5.7,\ am=0.025,\ N_f=2,\ 
  {\rm pion})
\eeq{pionmass}
which is stable against a variation of all the higher masses we included,
their prior confidence limits, the prior value of $m_0$, and the prior
confidence limit on it, as shown in \fgn{stable}. Exactly the same result
is obtained on changing the Bayesian analysis from MEM to MCF. This is
to be compared to the value $0.396\pm0.005$ reported by the simpler
analysis in \cite{ilgti}. In agreement with previous literature, we
find that Basyesian analysis can remove the contamination of masses by
higher-lying states which dogs the maximum likelihood analysis when the
local mass plateau is not fully developed. Our main new technical result
remains the observation of a cross over from prior to data dominated
regimes in the Bayesian analysis.

We end with a simple cautionary example: fitting a straight line through
a single point. This problem with $N_t<N_e$ is ill-posed without the
regularization provided by Bayesian priors. Let us say that we have
a measurement of $C(t)$ at $t=1$. The model is $C(t)=a+bt$, with the
parameters $a$ and $b$ to be determined. The $1\times2$ matrix $K$
is given by
\beq
   K=(1\;\;1) =(1) \; (\sqrt2\;\;0)\; \left(
    \begin{matrix}\frac1{\sqrt2} & \frac1{\sqrt2}\\-\,\frac1{\sqrt2} & \frac1{\sqrt2}\end{matrix}\right).
\eeq{kernel}
where we have also displayed its singular value decomposition (SVD). The
null space of $K$ is $a=-b$. If the measurement is $d\pm\sigma$, then
$\chi^2=(d-a-b)^2/\sigma^2$. Clearly $a=d$ and $b=0$ is a solution.
One can add to it any element from the null space of $K$, \ie, any
$a=d+z$ and $b=-z$ is also a solution. The quantity $z$ is completely
unconstrained. If we have priors $\rho^{(0)}_1=A$ and $\rho^{(0)}_2=B$,
then the MEM equations are
\beq
   d-a-b=\alpha\sigma^2\log\left(\frac aA\right), \qquad{\rm and}\qquad
   d-a-b=\alpha\sigma^2\log\left(\frac bB\right).
\eeq{memeq}
Now parametrizing $a=A\exp y$, one must also have $b=B\exp y$. The
search space is the line $a=(A/B)b$. Along this line the problem becomes
well posed,
\beq
   d-(A+B){\rm e}^y = \alpha\sigma^2 y.
\eeq{memfinal}
For $\alpha=0$ the solution is $a=Ad/(A+B)$ and $b=Bd/(A+B)$. For large
$\alpha$ the solution crosses over to the vicinity of the prior $a=A$
and $b=B$. Note that in neither region is the solution independent of
the prior.  This example shows that even after getting rid of $\alpha$,
the best fit parameters may still depend on priors. In this case the
Bayesian regulator allows the problem to be solved, but the solution
depends on the assumptions. One may immediately see that the problem is
structural, and does not depend on the kernel $K$. Since extraction of the
spectral function at finite temperature from Euclidean time correlators
is a mild generalization of this problem (larger $N_t$ and $N_e$ and a
different $K$), it behooves us to be extremely careful.

In this paper we have examined the Bayesian approach to parameter extraction
when all the parameters are not determined by the data. We have shown that
some nuisance parameters cause the solution to cross over from a prior
dominated to a data dominated region in the best of cases. When this happens,
then a reasonable way to deal with the nuisance parameters is to take them to
lie in the region where the solution is data driven. However, we have also
given an example of a problem where the priors never drop out of the solution.
As a result, one must always test the prior-dependence of the Bayesian fits.

Some of the numerical computations used lattice configurations and propagators
obtained using the computing resources of the Indian Lattice Gauge Theory
Initiative (ILGTI).

\end{document}

%% file: macros.tex
\newcommand\bef{\begin{figure}}
\newcommand\eef[1]{\label{fg:#1}\end{figure}}
\newcommand\beq{\begin{equation}}
\newcommand\eeq[1]{\label{#1}\end{equation}}
\newcommand\beqa{\begin{eqnarray}}
\newcommand\eeqa[1]{\label{#1}\end{eqnarray}}
\newcommand\bet{\begin{table}}
\newcommand\eet[1]{\label{tb:#1}\end{table}}

\newcommand\fgn[1]{Figure \ref{fg:#1}}
\newcommand\eqn[1]{eq.\ (\ref{#1})}

\newcommand\ie{{\sl i.e.\/}}

\newcommand\etal{{\sl et al.\/}}

\newcommand\pr{{\sl Phys.\ Rev.\/}\ }

